# Priming reasoning increases intentions to wear a face covering to slow down COVID-19 transmission

Valerio Capraro[1] and Hélène Barcelo[2]

[1] Middlesex University London, UK
[2] Mathematical Science Research Institute, Berkeley, USA

Contact author: v.capraro@mdx.ac.uk

**Abstract**

Finding mechanisms to promote the use of face masks is fundamental during the second phase of the COVID-19 pandemic response, when shelter-in-place rules are relaxed and some segments of the population are allowed to circulate more freely. Here we report three pre-registered studies (total N = 1,920), using an heterogenous sample of people living in the USA, showing that priming people to "rely on their reasoning" rather than to "rely on their emotions" significantly increases their intentions to wear a face covering. Compared to the baseline, priming reasoning promotes intentions to wear a face covering, whereas priming emotion has no significant effect. These findings have theoretical and practical implications. Practically, they offer a simple and scalable intervention to promote intentions to wear a face mask. Theoretically, they shed light on the cognitive basis of intentions to wear a face covering.

# Introduction

The coronavirus disease (COVID-19) pandemic is one of the greatest health threats of the last century. At the time of writing (June 11, 2020), over 7 millions of people have been tested positive and over 400,000 are confirmed dead[1] – and probably these are substantial underestimations (Burn-Murdoch, Romei, & Giles, 2020).

The large impact of COVID-19 is partly due to the fact that it can be transmitted by asymptomatic people, who often are not aware of being infected, through a cough or a sneeze (Bai et al. 2020; Mizumoto et al. 2020; Nishiura et al. 2020). For this reason, epidemiologists and health experts have recommended that people use face coverings, with the aim of minimizing the number of infected droplets that asymptomatic people spread around with the risk of infecting others. In line with these experts' suggestions, a study based in Germany found that the use of face mask reduced the daily growth rate of reported infections by about 40% (Mitze et al. 2020), whereas a study based in Beijing, China, exploring transmission in families with at least one laboratory confirmed COVID-19 case found that "face mask use by the primary case and family contacts before the primary case developed symptoms was 79% effective in reducing transmission" (Wang et al. 2020).

Yet, since wearing a face covering represents a significant change in people's habitual behaviour, we might expect that people would be reluctant to wear one. It follows that developing mechanisms that favour the use of face masks is crucial to slow down COVID-19 transmission and "flatten the curve" of the spread. Several national or local governments have taken the hard decision of making the use of face coverings mandatory in several contexts (Javid, 2020). However, since it is impossible to monitor the behavior of every person, even in places where wearing a face covering is mandatory, explicit laws should be complemented by implicit behavioral "nudges" aimed at directing people's behavior towards the desired one. In particular, appeals and messages can be very effective at promoting desired behavioral changes, because they can reach people inside their homes, through television and social media, as well as outside their homes, through screens, posters, and megaphones. This raises an important question. Which types of messages are most effective in promoting the use of face coverings?

Little is known about this question. Several papers have explored the effect of appeals and messages on intentions to engage in COVID-19 preventive behaviours (Barari et al. 2020; Bilancini et al. 2020; Capraro & Barcelo, 2020; Everett et al. 2020; Falco & Zaccagni, 2020; Heffner et al. 2020; Jordan et al. 2020; Lunn et al. 2020; Pfattheicher et al. 2020). However, with the exception of one paper, none of these works explored the effect of messages on intentions to wear a face covering; the only exception is Capraro and Barcelo (2020), which found that telling subjects that the coronavirus (COVID-19) is a threat to *their community* increases intentions to wear a face covering, relative to the baseline. This paper also showed a correlation such that people who feel negative emotions when wearing a face covering are less likely to self-report that they intend to wear a face covering. This suggests that deactivating the emotional cognitive system by asking people to rely on their reasoning, rather than on emotions, might be an effective tool to increase intentions to wear a face covering. So here we ask the following research question: Does priming people to rely on their reasoning, rather than to rely on their emotions, promote their intentions to use a face covering?

---
[1] https://www.worldometers.info/coronavirus/

This is obviously an important practical question: if it turns true, it would offer a simple scalable intervention to promote intentions to wear a face covering. But it is also an interesting theoretical question. According to dual-process theory, people's decisions result from the interplay between two cognitive processes, one that is fast, effortless and intuitive (named System 1) and one that is slow, effortful and deliberative (named System 2). (Fodor, 1983; Schneider & Schiffrin, 1977; Epstein & Pacini, 1999; Chen & Chaiken, 1999; Reber, 1993; Sloman, 1996; Kahneman & Frederick, 2005; Kahneman, 2011; Evans & Stanovich, 2013; De Neys & Pennycook, 2019). The dual-process approach has been shown useful to study people's decisions in several domains, including cooperation (Rand, 2016; Kvarven et al. 2020), altruism (Rand et al. 2016; Fromell et al. 2020), honesty (Köbis et al. 2019), reciprocity (Halsson et al. 2018), morality (Greene et al. 2008), and human-animal relationships (Caviola & Capraro, 2020). See Capraro (2019) for a detailed review. Emotion and reason are considered to be two "typical correlates" of System 1 and System 2, respectively (Evans & Stanovich, 2013). Therefore, exploring whether priming emotion vs. reason affects people's intentions to wear a face covering would allow us to shed light on the cognitive basis of intentions to wear a face mask.

In this paper, we report three pre-registered experiments (total N=1,920). The experiments have been conducted on a heterogeneous, although not representative, sample of people living in the USA and surveyed using Amazon Mechanical Turk (Paolacci, Chandler, & Ipeirotis, 2010). We found that priming people to rely on their reasoning increases intentions to wear a face covering relative to the "symmetric" priming, according to which people are primed to rely on their emotions. Comparing the primes to the baseline, we find that the effect is primarily driven by reasoning, meaning that, compared to the baseline, priming reasoning significantly increases intentions to wear a face covering, whereas priming emotions do not significantly change intentions to wear a face covering.

## Study 1

**Method**

*Conditions*

Participants were randomly assigned to one of two conditions: in the *priming emotion* condition, they were shown a message highlighting the positive consequences of making decisions based on feelings; in the *priming reasoning* condition, they were shown a message highlighting the positive consequences of making decisions based on reasoning. These primes were taken from previously published work (Levine et al. 2018; Capraro, Everett & Earp, 2019; Caviola & Capraro, 2020). See Table 1 for the exact primes.

| Condition | Message |
|---|---|
| Priming emotion | Sometimes people make decisions by using feelings and relying on their emotions. Other times, people make decisions by using logic and relying on their reasoning. Many people believe that emotions lead to good decision-making. When we use feelings, rather than logic, we make emotionally satisfying decisions. Please |

| | answer the following questions by relying on emotions, rather than reasoning. |
|---|---|
| Priming reason | Sometimes people make decisions by using logic and relying on their reasoning. Other times, people make decisions by using feelings and relying on their emotions. Many people believe that reason leads to good decision-making. When we use logic, rather than feelings, we make rationally satisfying decisions. Please answer the following questions by relying on reasoning, rather than emotions. |

*Table 1. Conditions of the experiment. Between-subjects random assignment.*

### Dependent variables

After reading the prime, all participants took the following scale.

> *Intentions to wear a face covering.* Participants were asked to: "answer the following questions by relying on emotions [reasoning]. When the shelter-in-place rules are relaxed, I intend to ...
> a. Wear a face covering any time I leave home.
> b. Wear a face covering any time I am engaged in essential activities and/or work, and there is no substitute for physical distancing and staying at home.
> c. Wear a face covering any time I'm around people outside my household."

All answers were collected using a 10-line "snap to grid" slider with three labels: "strongly disagree" at the extreme left, "neither agree nor disagree" at the center, "strongly agree" at the extreme right.

### Demographics

After the scale, participants were asked the following set of demographic questions: sex, age, race, political views, religiosity, whether they live in an urban area, whether wearing a face covering is mandatory in their county, whether there live in an area where shelter-in-place rules apply, whether they were tested positive, whether they believe they will get the coronavirus and, if so, whether they believe they will get over it relatively easily". At the end, there was a control question to get rid of potential bots.

### Pre-registration

The design and the analyses were pre-registered at: https://aspredicted.org/h3x5d.pdf.

## Results

### Demographic characteristics of the sample

As pre-registered, we eliminated from the analysis subjects who did not pass the attention check and, for each multiple IP address or Turk ID, we keep only the first observation and discarded the rest. This corresponds to deleting about 2% of the observations. Our main results remain qualitatively the same when including these observations. We report in Table A1 the demographic characteristics of the sample for this and the following studies. We note that the sample is quite heterogeneous, although not representative: males and females are equally represented; the age group 25-54 is overrepresented, whereas the age groups 18-24 and 65+ are underrepresented; Whites are overrepresented, while Blacks or African Americans are underrepresented.[2] The average participant is more left-leaning than the average American.[3] The sample is representative of people living in urban areas.[4] We could not find information about the exact percentage of counties where wearing a face covering was mandatory and about the percentage of counties where shelter-in-place rules applies; in Table A1 we report also these percentages for completeness.

*The effect of priming emotion vs. reasoning on intentions to wear a face covering*

We first build the composite variable "intentions to wear a face covering", by taking the average of its three items ($\alpha_{emotion}$ = 0.932, $\alpha_{reason}$ = 0.924). Wilcoxon rank-sum shows that the distribution of intentions to wear a face covering when reasoning is primed is statistically different from the one when emotion is primed ($z$ = 2.366, $p$ = 0.018). Looking at the averages, promoting reasoning appears to increase intentions to wear a face covering relative to priming emotion ($M_{reason}$ = 7.38, $SD_{reason}$ = 3.00; $M_{emotion}$ = 6.61, $SD_{emotion}$ = 3.24).

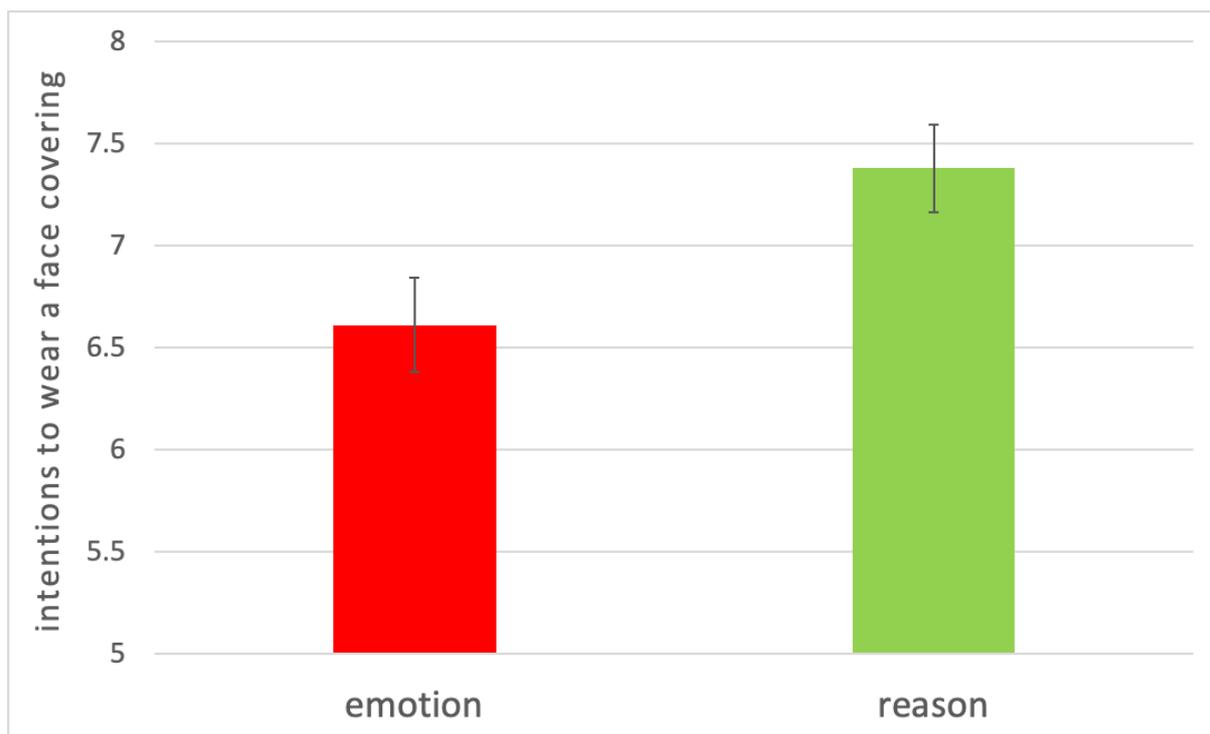

*Figure 1. Intentions to wear a face covering split by treatment; y-axis from 0 to 10. Error bars represent the standard error of the mean.*

---

[2] https://en.wikipedia.org/wiki/Demographics_of_the_United_States
[3] https://news.gallup.com/poll/275792/remained-center-right-ideologically-2019.aspx
[4] https://www.census.gov/newsroom/press-releases/2016/cb16-210.html

# Study 2

Study 1 shows that priming reasoning vs emotion promotes intentions to wear a face covering. However, it is not clear whether it is priming reasoning that promotes intentions to wear a face covering, or it is priming emotion that undermines intentions to wear a face covering, or both. To answer this question, in Study 2 we repeat the experiment by adding a baseline condition. Apart from answering our main question, this would also be an occasion to replicate the results of Study 1 (Open Science Collaboration, 2015).

**Method**

*Conditions*

Study 2 is identical to Study 1 with the only difference that we added the baseline, so that participants in Study 2 are randomly divided in three conditions: *priming emotion*, *baseline*, and *priming reasoning*.

*Pre-registration*

The design, the analyses, and the sample size were pre-registered at: https://aspredicted.org/bq9v6.pdf.

**Results**

*Demographic characteristics of the sample*

People who participated in the previous study, as well as those who participated in the study reported in Capraro & Barcelo (2020), have not been allowed to participate in this study. The demographic characteristics of the sample are reported in Appendix, Table A1.

*The effect of priming emotion vs reason on intentions to wear a face covering*

We first build the composite variable "intentions to wear a face covering", by taking the average of its three items ($\alpha_{emotion}$ = 0.914, $\alpha_{baseline}$ = 0.937, $\alpha_{reason}$ = 0.941). A one-way ANOVA with Bonferroni correction[5] reveals that there were not statistically significant differences across conditions ($F(2,588) = 0.31$, $p = 0.731$). See Figure 2.

---

[5] We pre-registered that we would use pairwise rank-sum, but then we realized that a one-way ANOVA with Bonferroni correction would be the correct test to be used in this case. In any case, we note that the pairwise rank-sum tests, after Bonferroni correction, give qualitatively the same results as the post-hoc tests of the ANOVA.

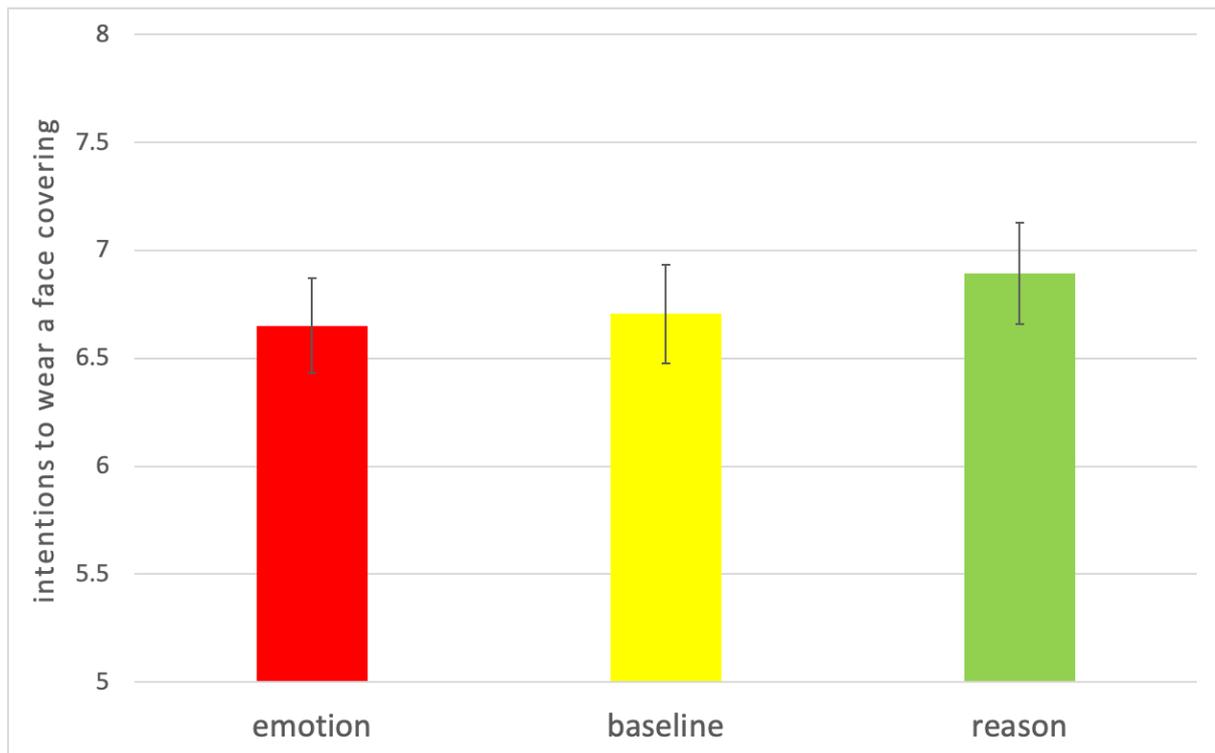

*Figure 2. Intentions to wear a face covering split by treatment in Study 2; y-axis from 0 to 10. Error bars represent the standard error of the mean.*

## Study 3

Study 2 finds a non-significant trend in the same direction as Study 1. One possibility is that Study 1 was a false positive. Another possibility is that Study 2 failed to find an effect for some reasons (e.g., Study 2 was conducted the Monday after the protests against George Floyd's death, and this might have pushed some people to make decisions using their emotion, thus diminishing the effect of reasoning). To clarify this, we conducted a third study with a larger sample size sufficient to detect a small effect of $d = 0.20$ with power 0.80 and alpha = 0.05.

**Method**

*Conditions*

Study 3 is identical to Study 2.

*Pre-registration*

The design, the analyses, and the sample size were pre-registered at: https://aspredicted.org/qb5y8.pdf.

**Results**

*Demographic characteristics of the sample*

People who participated in the previous two studies, as well as those who participated in the study reported in Capraro & Barcelo (2020), have not been allowed to participate in this study. The demographic characteristics of the sample are reported in the Appendix, Table A1.

*The effect of priming emotion vs reason on intentions to wear a face covering*

We first build the composite variable "intentions to wear a face covering", by taking the average of its three items ($\alpha_{emotion}$ = 0.933, $\alpha_{baseline}$ = 0.941, $\alpha_{reason}$ = 0.928). A one-way ANOVA with Bonferroni correction reveals a statistically significant effect of condition on intentions to wear a face covering ($F(2,927) = 7.35$, $p < 0.001$). Post-hoc comparisons show that intentions to wear a face covering when reasoning is primed ($M = 7.23$, $SD = 2.97$) are significantly higher than intentions to wear a face covering when emotion is primed ($M = 6.29$, $SD = 3.10$), $p < .001$. On the other hand, intentions to wear a face covering in the baseline ($M = 6.71$, $SD = 3.22$) does not appear to be significantly different than the other two conditions (both $p$'s > 0.12). See Figure 3.

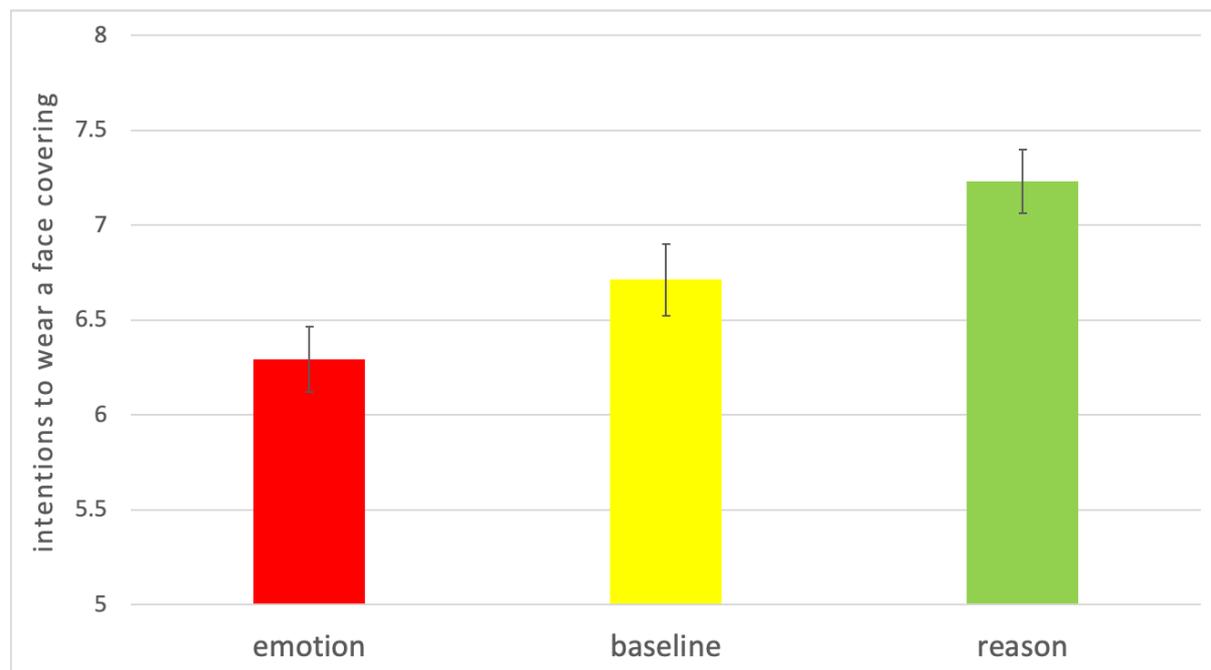

*Figure 3. Intentions to wear a face covering split by treatment in Study 3; y-axis from 0 to 10. Error bars represent the standard error of the mean. Note that the errors do not take into account Bonferroni correction.*

## All studies together

As pre-registered in Study 3, we now put all the data together to increase the power and test which of the three effects are significant with a larger sample size. A one-way ANOVA with Bonferroni correction confirms the statistically significant effect of condition ($F(2,1917) = 9.17$, $p < .001$). Post-hoc comparisons reveal that intentions to wear a face covering are higher in the *priming reason* condition ($M = 7.18$, $SD = 3.09$) compared to the *priming emotion* condition ($M = 6.48$, $SD = 3.12$), $p < .001$, whereas intentions to wear a face covering in the baseline ($M = 6.71$, $SD = 3.21$) lays between the two other conditions, but it is significantly different from the *priming reason* condition (*p* = .033) but not from the *priming emotion* condition (*p* = 0.614). See Figure 4.

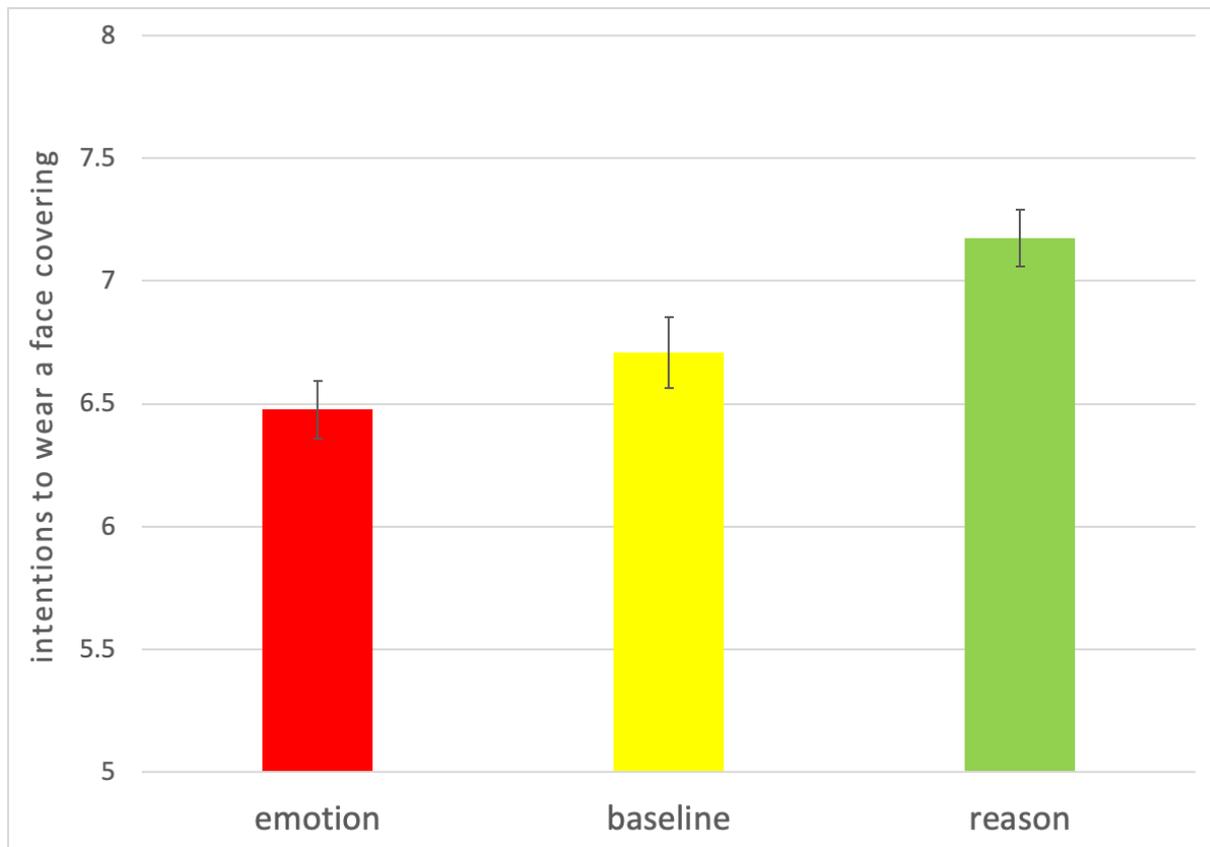

*Figure 4. Intentions to wear a face covering split by treatment, all studies together; y-axis from 0 to 10. Error bars represent the standard error of the mean. Note that the errors do not take into account Bonferroni correction.*

### *Exploratory analysis looking at potential moderators of the effect*

As exploratory analysis, we add each demographic variable as a separated moderator, in order to test whether the effect of the treatment is particularly strong on a subset of participants. In doing so, we find no significant moderation (sex: p = 0.896, age: p = 0.430; race: p = 0.602; political views: p = 0.164; religiosity: p = 0.735; living in urban area: p = 0.104; living in a county where wearing a face covering is mandatory: p = 0.239; living in a county where there are shelter-in-place rules: p = 0.262; tested positive: p = 0.865; tested negative: p = 0.594). This suggests that the effect of priming reason vs. emotion is relatively stable across subsets of the population.

## Discussion

Here we reported three pre-registered studies exploring the effect of priming emotion vs reasoning on intentions to wear a face covering. Altogether, these studies show that priming reasoning increases intentions to wear a face covering relative to priming emotions. Compared to the baseline, priming reasoning increases intentions to wear a face covering, whereas priming emotion has no effect on intentions to wear a face covering. The latter finding should be taken with caution, because the data trend in the direction that priming emotion decreases intentions to wear a face covering, compared to baseline. Therefore, it is possible that we failed to detect the effect of priming emotion vs. baseline because of insufficient statistical power.

These results have practical and theoretical implications. From a practical perspective, finding ways to promote the use of face masks is key during the second phase of the COVID-19 pandemic response, in which, after the initial strict lockdown, local and national governments are relaxing shelter-in-place rules, so that some segments of the population are allowed to circulate more freely. Since some of these people will be positive to COVID-19 without being aware of it, wearing a face mask helps to decrease the probability that infected droplets are spread around and infect other people. In this light, our results suggest a simple and scalable intervention to promote *intentions* to use face masks: priming people to rely on their reasoning. Of course, our results have some limitations. One regards the sample. Our results were obtained with a heterogeneous, but not representative, sample of people living in the US. However, we included each demographic variable as a potential moderator into separate regression models. In doing so, we found that none of the demographic variables moderated the effect of the primes on intentions to wear a face covering. This provides a piece of evidence that our results can be generalized to the American population at large. Of course, our results cannot be readily generalized to other countries. Future work should test the effect of conceptual primes on intentions to wear a face mask in different countries. A major limitation of our study is instead the fact that it focuses on intentions, rather than actual behavior. A recent study found that intentions to practice physical distancing is correlated to actual behavior (Gollwitzer et al. 2020). Although this certainly does not imply that intentions to wear a face covering correlate with actual behavior, it does give some hope that it will actually be the case. Future work should test whether conceptual primes of the form used in this paper impact people's actual use of face coverings.

From a theoretical perspective, our results suggest that intentions to wear a face covering are affected by the cognitive system that is primarily active during the decision-making process. This because relying on emotions is typically correlated with System 1 processing, whereas relying on reasoning is typically correlated with System 2 processing (Evans & Stanovich, 2020). This suggests other potential techniques to promote the use of face coverings tapping to the same cognitive processes. Indeed, conceptual primes of emotion and reason are not the only experimental techniques used to activate System 1 and System 2 respectively. Other techniques include time constraints, ego depletion and cognitive load (see Capraro, 2019, for a review). It is possible that some of these techniques could be even more effective at promoting the use of face masks than conceptual primes. Future work could therefore test which cognitive manipulations work best at promoting the use of face masks.

# Appendix

**Demographic characteristics of the sample**

| Demographic | | Percent | | | |
|---|---|---|---|---|---|
| | | **Study 1** (N=399) | **Study 2** (N=591) | **Study 3** (N=930) | **All studies** (N=1,920) |
| gender | female | 50.63 | 50.59 | 51.29 | 50.93 |
| | male | 48.62 | 49.24 | 48.38 | 48.70 |
| | prefer not to say | 0.75 | 0.17 | 0.32 | 0.37 |
| age | 18-24 | 9.27 | 7.95 | 6.77 | 7.65 |
| | 25-34 | 34.59 | 35.70 | 40.32 | 37.71 |
| | 35-44 | 27.32 | 29.27 | 25.37 | 26.98 |
| | 45-54 | 14.04 | 16.92 | 14.30 | 15.05 |
| | 55-64 | 9.77 | 6.93 | 8.82 | 8.43 |
| | 65+ | 4.51 | 3.21 | 4.41 | 4.11 |
| race | American Indian or Alaska native | 1.00 | 0.51 | 0.97 | 0.83 |
| | Asian | 11.02 | 8.13 | 9.82 | 9.56 |
| | Black or African American | 6.77 | 7.11 | 9.06 | 7.99 |
| | Native Hawaiian or other Pacific Islander | 0 | 0 | 0 | 0 |
| | White | 77.19 | 80.33 | 75.72 | 77.49 |
| | Multiracial | 3.76 | 3.89 | 4.42 | 4.13 |
| political view | left-leaning | 47.61 | 46.36 | 46.01 | 46.67 |
| | center | 25.56 | 23.18 | 25.53 | 24.79 |
| | right-leaning | 26.82 | 29.61 | 29.46 | 28.54 |
| living in an urban or suburban area | | 81.70 | 78.17 | 78.04 | 78.84 |
| living in a county where wearing a face covering is mandatory | | 49.12 | 45.01 | 44.40 | 45.57 |
| living in a county where there are shelter in place rules | | 64.16 | 56.51 | 50.21 | 55.05 |

*Table A1. Demographic characteristics of the overall sample. Political view goes from 1 = "very left-leaning" to 7 = "very right-leaning", with 4 = "center". In the Table we classified as "center" only those subjects who answered "center".*